\pdfoutput=1
\documentclass[11pt, a4paper]{lumia}
\usepackage[sort&compress]{natbib}
\usepackage{longtable}   
\usepackage{array}      
\usepackage{xltabular}
\usepackage{arydshln}
\usepackage{dashrule}
\usepackage{booktabs}
\usepackage{caption}

\usepackage{booktabs}
\usepackage{multirow}
\usepackage{graphicx}
\usepackage{tabularx}
\usepackage{array}
\usepackage{pifont}
\usepackage[table]{xcolor}
\newcommand{\leaderlogo}[1]{%
  \raisebox{-0.22ex}{\includegraphics[height=2ex,keepaspectratio]{#1}}%
}
\usepackage{wrapfig}
\usepackage{xcolor}
\newcommand{\upperformence}[1]{\textcolor{green!50!black}{+#1}}
\newcommand{\downperformece}[1]{\textcolor{red!70!black}{-#1}}

\usepackage[most]{tcolorbox}

\setheadertext{PatRe: A Full-Stage Office Action and Rebuttal Generation Benchmark for Patent Examination}

\renewcommand{\today}{2026-10}

\usepackage{xspace}

\theoremstyle{plain}

\newtheorem*{proposition*}{Proposition}

\theoremstyle{definition}

\theoremstyle{definition}

\def\eqref#1{equation~\ref{#1}}

\usepackage[utf8]{inputenc}     
\usepackage[T1]{fontenc}        
\usepackage{microtype}          

\usepackage{xcolor}             
\usepackage[dvipsnames]{xcolor} 
\usepackage{graphicx}           
\usepackage{tikz}               
\usepackage[edges]{forest}      

\usepackage{hyperref}           
\usepackage{url}                
\usepackage{xurl}               

\usepackage{tabularx}           
\usepackage{array}              
\usepackage{booktabs}           
\usepackage{longtable}          
\usepackage{multirow}           
\usepackage{makecell}           
\usepackage[table]{xcolor}      
\usepackage{colortbl}           
\usepackage{ragged2e}           

\usepackage{amsmath}            
\usepackage{mathtools}          
\usepackage{amsfonts}           
\usepackage{amssymb}            
\usepackage{amsthm}             
\usepackage{nicefrac}           

\usepackage{algorithm}          
\usepackage{algorithmicx}       
\usepackage{algpseudocode}      
\usepackage{listings}           

\usepackage{subcaption}         
\usepackage{caption}            
\usepackage{wrapfig}            
\usepackage[export]{adjustbox}  

\usepackage{soul}               
\usepackage[commandnameprefix=always]{changes}            
\usepackage{xspace}             
\usepackage[normalem]{ulem}     
\usepackage{CJKutf8}            

\usepackage{enumitem}           
\usepackage{pifont}             

\usepackage[tikz]{bclogo}       
\usepackage[framemethod=tikz]{mdframed} 

\usepackage{lipsum}             
\usepackage{tocloft}            
\usepackage{fontawesome5}       
\usepackage{afterpage}          
\usepackage{bbding}             
\usepackage{epigraph}           
\usepackage{minitoc}            
\usepackage{multicol}           
\usepackage{textgreek}          

\newcolumntype{P}[1]{>{\RaggedRight\arraybackslash}p{#1}}

\definecolor{darkblue}{rgb}{0, 0, 0.5}
\definecolor{uclablue}{RGB}{39, 116, 174}
\definecolor{bigaired}{RGB}{156, 0, 0}
\definecolor{myblue}{HTML}{598BE7}
\definecolor{mildblue}{RGB}{31,119,180}
\definecolor{sectionblue}{RGB}{70, 130, 180}
\definecolor{methodblue}{RGB}{0, 150, 136}
\definecolor{bgblue}{RGB}{245,243,253}
\definecolor{ttblue}{RGB}{91,194,224}
\definecolor{mygreen}{rgb}{0.64, 0.56, 0.88}
\definecolor{myyellow}{rgb}{0.68, 0.6, 0.1}
\definecolor{fancygreen}{rgb}{0.33, 0.68, 0.20}
\definecolor{salmon}{rgb}{0.94, 0.52, 0.49}
\definecolor{tablegreen}{rgb}{0.82, 0.94, 0.75}
\definecolor{tableblue}{rgb}{0.81, 0.90, 0.94}
\definecolor{tablered}{rgb}{0.97, 0.85, 0.85}
\definecolor{tableorange}{rgb}{0.96, 0.85, 0.81}
\definecolor{myorange}{rgb}{1.0, 0.49, 0.0}
\definecolor{tlgreen}{rgb}{0.33, 0.68, 0.20}
\definecolor{darkgreen}{RGB}{0,100,0}
\definecolor{darkred}{RGB}{200, 0, 0}
\definecolor{lightblue}{RGB}{220,235,250}
\definecolor{customyellow}{HTML}{FFFACD}
\definecolor{refinegreen}{RGB}{0, 128, 75}
\definecolor{scoregreen}{RGB}{34, 139, 34}
\definecolor{hidden-blue}{RGB}{194,232,247}
\definecolor{hidden-black}{RGB}{20,68,106}
\definecolor{yes}{HTML}{C6EFCE}
\definecolor{no}{HTML}{FFC7CE}
\definecolor{partial}{HTML}{FFEB9C}
\definecolor{external}{HTML}{D9E1F2}
\definecolor{hdr}{HTML}{F2F2F2}
\definecolor{GRPOrow}{gray}{0.96}
\definecolor{FlowRLrow}{RGB}{225,236,255}
\definecolor{FlowBlue}{RGB}{80,120,210}
\definecolor{GRPOGray}{gray}{0.35}

\hypersetup{
    colorlinks=true, 
    citecolor=uclablue, 
    linkcolor=bigaired,
    urlcolor=darkblue
}

\setlist[itemize]{leftmargin=20pt, noitemsep, topsep=0pt}

\NewDocumentCommand{\kaiyan}{mO{}}{\textcolor{purple}{\textsuperscript{\textit{kaiyan}}\textsf{\textbf{\small[#1]}}}}
\NewDocumentCommand{\yuxin}{mO{}}{\textcolor{cyan}{\textsuperscript{\textit{yuxin}}\textsf{\textbf{\small[#1]}}}}
\NewDocumentCommand{\bx}{mO{}}{\textcolor{green}{\textsuperscript{\textit{bx}}\textsf{\textbf{\small[#1]}}}}
\NewDocumentCommand{\at}{mO{}}{\textcolor{red}{\textsuperscript{\textit{AT}}\textsf{\textbf{\small[#1]}}}}
\NewDocumentCommand{\re}{mO{}}{\textcolor{blue}{\textsuperscript{\textit{RE}}\textsf{\textbf{\small[#1]}}}}
\NewDocumentCommand{\ybsun}{mO{}}{\textcolor{magenta}{\textsuperscript{\textit{youbang}}\textsf{\textbf{\small[#1]}}}}
\NewDocumentCommand{\runze}{mO{}}{\textcolor{orange}{\textsuperscript{\textit{runze}}\textsf{\textbf{\small[#1]}}}}
\NewDocumentCommand{\add}{mO{}}{\textcolor{darkgreen}{\textsuperscript{\textit{Maybe Consider Discuss}}\textsf{\textbf{[#1]}}}}

\newcommand{\cmark}{\textcolor{darkgreen}{\boldmath$\checkmark$}}
\newcommand{\xmark}{\textcolor{darkred}{\boldmath$\times$}}

\newenvironment{itemize*}%
 {\leftmargini=10pt\begin{itemize}%
  \setlength{\itemsep}{0pt}%
  \setlength{\parskip}{0pt}%
  }%
 {\end{itemize}}

\newenvironment{enumerate*}%
 {\begin{enumerate}%
  \setlength{\itemsep}{0pt}%
  \setlength{\parskip}{0pt}}%
 {\end{enumerate}}

\newcommand{\cellstatus}[1]{%
  \begingroup
  \StrTrim{#1}[\statusval]%
  \IfStrEq{\statusval}{Yes}{\cellcolor{yes}\cmark}{}%
  \IfStrEq{\statusval}{No}{\cellcolor{no}\xmark}{}%
  \IfBeginWith{\statusval}{Yes (}{\cellcolor{yes}\cmark~\textit{\statusval\unskip}}{}%
  \IfStrEq{\statusval}{Partial}{\cellcolor{partial}\textbf{Partial}}{}%
  \IfStrEq{\statusval}{External}{\cellcolor{external}\textbf{External}}{}%
  \endgroup
}

\newtcolorbox{myboxi}[1][]{
  breakable,
  title=#1,
  colback=red!5,
  colbacktitle=red!5,
  coltitle=black,
  fonttitle=\bfseries,
  bottomrule=0pt,
  toprule=0pt,
  leftrule=2pt,
  rightrule=2pt,
  titlerule=0pt,
  arc=0pt,
  outer arc=0pt,
  colframe=red,
}

\newtcolorbox{myboxnote}[1][]{
  breakable,
  title=#1,
  colback=orange!0,
  colbacktitle=orange!0,
  coltitle=black,
  fonttitle=\bfseries,
  bottomrule=0pt,
  toprule=0pt,
  leftrule=2pt,
  rightrule=2pt,
  titlerule=0pt,
  arc=0pt,
  outer arc=0pt,
  colframe=orange,
}

\newtcolorbox{myboxii}[1][]{
  breakable,
  freelance,
  title=#1,
  colback=white,
  colbacktitle=white,
  coltitle=black,
  fonttitle=\bfseries,
  bottomrule=0pt,
  boxrule=0pt,
  colframe=white,
  overlay unbroken and first={
  \draw[red!75!black,line width=3pt]
    ([xshift=5pt]frame.north west) -- 
    (frame.north west) -- 
    (frame.south west);
  \draw[red!75!black,line width=3pt]
    ([xshift=-5pt]frame.north east) -- 
    (frame.north east) -- 
    (frame.south east);
  },
  overlay unbroken app={
  \draw[red!75!black,line width=3pt,line cap=rect]
    (frame.south west) -- 
    ([xshift=5pt]frame.south west);
  \draw[red!75!black,line width=3pt,line cap=rect]
    (frame.south east) -- 
    ([xshift=-5pt]frame.south east);
  },
  overlay middle and last={
  \draw[red!75!black,line width=3pt]
    (frame.north west) -- 
    (frame.south west);
  \draw[red!75!black,line width=3pt]
    (frame.north east) -- 
    (frame.south east);
  },
  overlay last app={
  \draw[red!75!black,line width=3pt,line cap=rect]
    (frame.south west) --
    ([xshift=5pt]frame.south west);
  \draw[red!75!black,line width=3pt,line cap=rect]
    (frame.south east) --
    ([xshift=-5pt]frame.south east);
  },
}

\tcbset{
  takeawaysbox/.style={
    title=Takeaways,
    colback=lightblue!80,
    colframe=black,
    fonttitle=\bfseries\small,
    coltitle=white,
    colbacktitle=black,
    enhanced,
    attach boxed title to top left={xshift=2.5mm,yshift=-2.5mm},
    boxed title style={rounded corners, size=small, colframe=black, colback=black},
    width=\linewidth,
    arc=3.5mm
  }
}

\mdfdefinestyle{mystyle}{%
  rightline=true,
  innerleftmargin=10,
  innerrightmargin=10,
  outerlinewidth=3pt,
  topline=false,
  rightline=true,
  bottomline=false,
  skipabove=\topsep,
  skipbelow=\topsep
}

\tikzset{%
    every node/.style={font=\tiny},
    parent/.style =          {align=center,text width=2cm,rounded corners=3pt, line width=0.3mm, fill=gray!10,draw=gray!80},
    child/.style =           {align=center,text width=2.0cm,rounded corners=3pt, fill=blue!10,draw=blue!80,line width=0.3mm},
    grandchild/.style =      {align=center,text width=2cm,rounded corners=3pt},
    greatgrandchild/.style = {align=center,text width=1.5cm,rounded corners=3pt},
    greatgrandchild2/.style = {align=center,text width=1.5cm,rounded corners=3pt},    
    referenceblock/.style =  {align=center,text width=1.5cm,rounded corners=2pt},
    pretrain/.style =           {align=center,text width=2.0cm,rounded corners=3pt, fill=blue!10,draw=blue!80,line width=0.3mm},   
    pretrain_work/.style =           {align=center, text width=8.5cm,rounded corners=3pt, fill=blue!10,draw=blue!0,line width=0.3mm},  
    template/.style =           {align=center,text width=2.0cm,rounded corners=3pt, fill=red!10,draw=red!80,line width=0.3mm},   
    template_work/.style =           {align=center,text width=8.5cm,rounded corners=3pt, fill=red!10,draw=red!0,line width=0.3mm},    
    answer/.style =           {align=center,text width=2.0cm,rounded corners=3pt, fill= cyan!10,draw= cyan!80,line width=0.3mm},   
    answer_work/.style =           {align=center,text width=8.5cm,rounded corners=3pt, fill= cyan!10,draw= cyan!0,line width=0.3mm},      
    multiple/.style =           {align=center,text width=2.0cm,rounded corners=3pt, fill= orange!10,draw= orange!80,line width=0.3mm},   
    multiple_work/.style =           {align=center,text width=8.5cm,rounded corners=3pt, fill= orange!10,draw= orange!0,line width=0.3mm},        
    tuning/.style =           {align=center,text width=2.0cm,rounded corners=3pt, fill= magenta!10,draw= magenta!80,line width=0.3mm},   
    tuning_work/.style =           {align=center,text width=8.5cm,rounded corners=3pt, fill= magenta!10,draw= magenta!0,line width=0.3mm},          
}

\newcommand{\lstbg}[3][0pt]{{\fboxsep#1\colorbox{#2}{\strut #3}}}

\lstdefinelanguage{diff}{
  basicstyle=\ttfamily\small,
  morecomment=[f][\lstbg{red!20}]-,
  morecomment=[f][\lstbg{green!20}]+,
}

\lstdefinelanguage{diffpython}{
  language=diff,
  morekeywords={def, if, else, for, while, return, import, from, as, class, with, try, except, finally, raise, lambda, and, or, not, in, is, None, True, False},
  morecomment=[l]{\#},
  morestring=[b]",
  morestring=[b]',
}

\newtcolorbox{promptbox}[2][]{
  colback=gray!5,       
  colframe=black!70,    
  title=\textbf{#2},     
  fonttitle=\bfseries\small,
  fontupper=\footnotesize\ttfamily, 
  left=3pt, right=3pt, top=3pt, bottom=3pt, 
  boxrule=0.8pt,
  sharp corners,         
  #1
}

\title{Novel or Not? PatEx-Gym: Agentic Prior-Art Search for Novelty Assessment in Patent Examination}

\author{%
    \scriptsize
    Qiyao Wang$^{1,2}$, Xinyi Chen$^{3}$, 
    Hongbo Wang$^{1,3}$,
    Jiayuan Lu$^{3}$,
    Jianhang Li$^{3}$,
    Minghuan Tan$^{1}$,
    Yuan Lin$^{3\dagger}$, Min Yang$^{1,4\dagger}$ \\
    $^1$~\raisebox{-0.5ex}{\includegraphics[height=2.9ex]{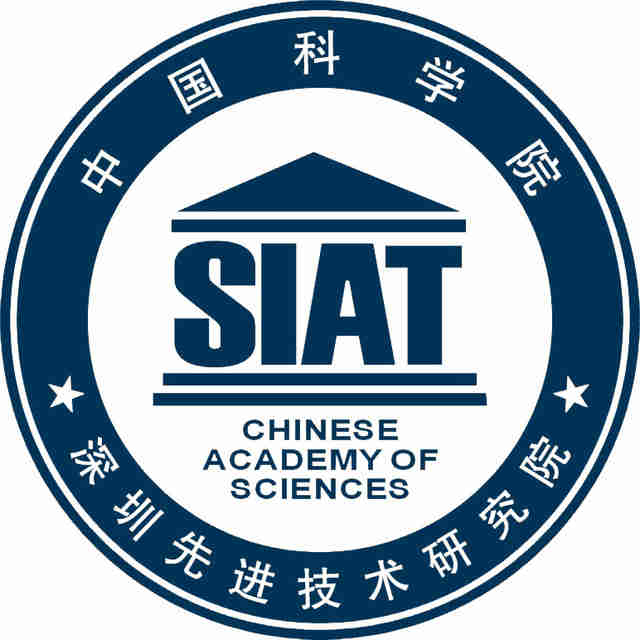}}~Shenzhen Institute of Advanced Technology, Chinese Academy of Sciences \quad
    $^2$~\raisebox{-0.5ex}{\includegraphics[height=2.9ex]{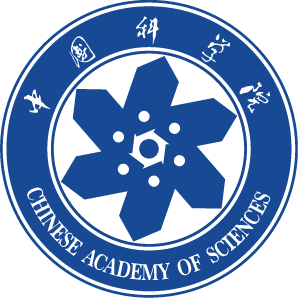}}~University of Chinese Academy of Sciences\\
    $^3$~\raisebox{-0.5ex}{\includegraphics[height=2.9ex]{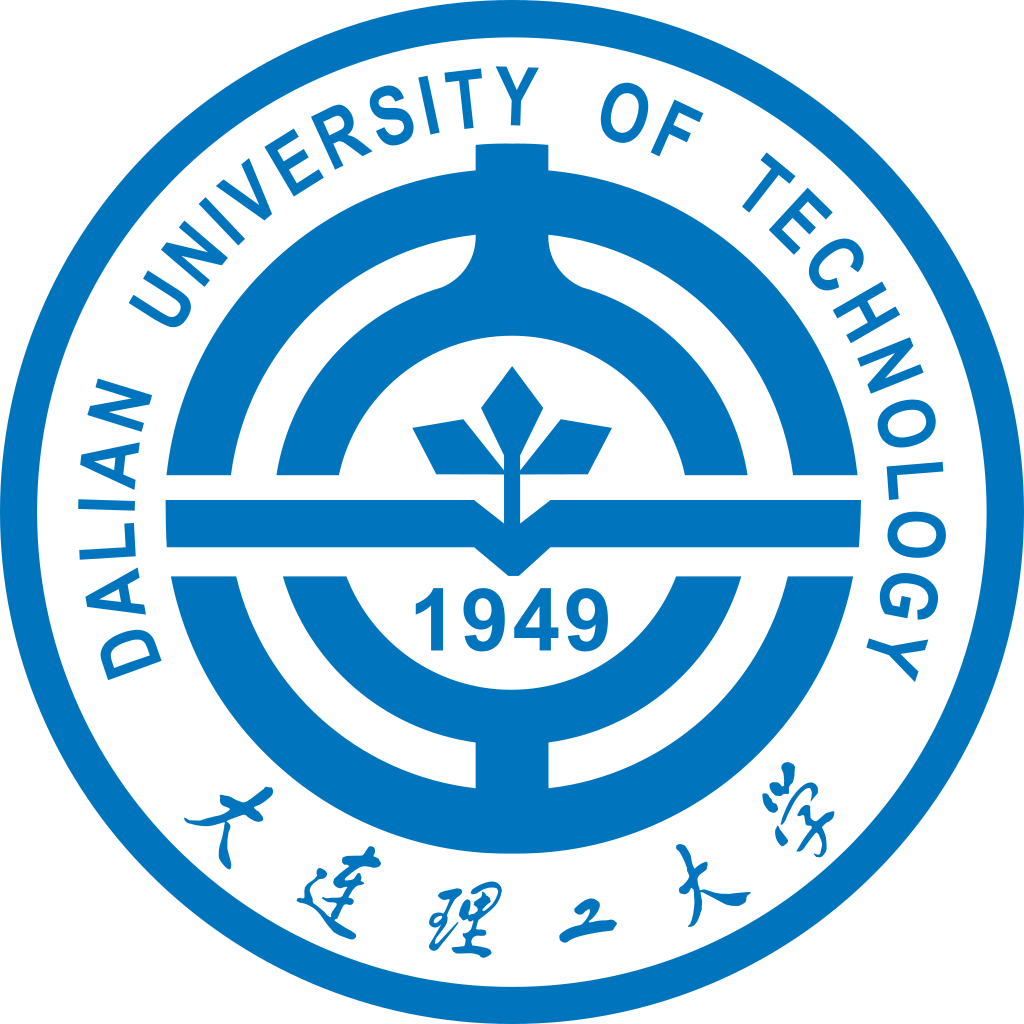}}~Dalian University of Technology\quad
    $^4$~\raisebox{-0.5ex}{\includegraphics[height=2.9ex]{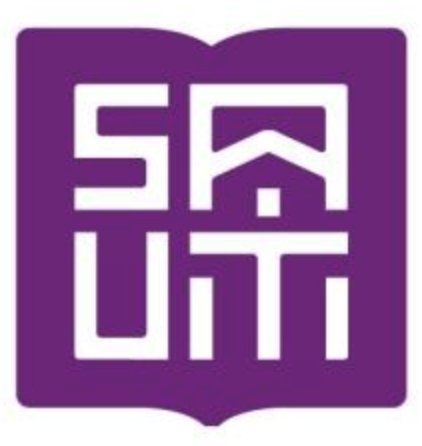}}~Shenzhen University of Advanced Technology \\
    \faEnvelope[regular]~\texttt{wangqiyao25@mails.ucas.ac.cn}  \quad
    \faEnvelope[regular]~\texttt{zhlin@dlut.edu.cn}  \quad
    \faEnvelope[regular]~\texttt{min.yang@siat.ac.cn}  \\
    \faHome~\href{https://patex-gym.wangqiyao.me/}{Website} \quad \faGithub~\href{https://github.com/AIforIP/PatEx-Gym}{PatRe} \quad
    $^*$ Equal Contribution. \quad
    $^\dagger$ Corresponding Authors. 
}

\begin{abstract}
Patent examination is a challenging real-world task that requires agents to discover sparse prior art, iteratively refine queries, inspect technical evidence, and produce novelty assessments documented in Office Actions~(OAs) across heterogeneous scientific and technical domains.
 However, existing benchmarks largely focus on static and limited candidate sets, leaving the full search--reasoning--assessment process underexplored.
 We present \textbf{PatEx-Gym}, the first framework for developing and evaluating agents for patent examination, which integrates three components. 
 First, \textbf{PatEx-Bench} provides 1K realistic examination tasks, two large-scale retrieval corpora containing over 1M patents and 52K scientific papers, and a harness runtime supporting iterative search, document visits, reasoning, and OA generation. 
 Second, we introduce \textbf{PASS}, an annotation-free {\underline{P}}rior-{\underline{A}}rt Agentic {\underline{S}}earch Trajectory {\underline{S}}ynthesis framework. 
 Specifically, it exploits document visits as implicit relevance signals and evidence-acquisition anchors to model intra-turn query behavior, inter-turn search progression, and trajectory-level evidence and outcome quality, yielding \textbf{PatEx-SynData}. 
 Third, we train \textbf{PatEx-Agent} at the 4B scale via supervised fine-tuning on the curated trajectories, establishing a lightweight baseline.
 Extensive experiments show that final decision accuracy alone does not fully reflect examination quality, while prior-art discovery remains difficult even for strong frontier models. 
 Our PatEx-Agent achieves competitive performance against substantially larger general-purpose agents. 
 These results show that training on PatEx-SynData enables a compact 4B model to serve as a competitive baseline for prior-art search and patent examination.
 We view patent examination as a demanding testbed for agentic search and deep research agents.
\end{abstract}

\begin{document}

 \maketitle

\section{Introduction}

Information retrieval systems have long served as a fundamental gateway to information access, spanning general-purpose Web search~\citep{brin-web-98,burges-l2r-05} and domain-specific search engines in knowledge-intensive fields such as law~\citep{feng-etal-2024-legal,harveylab2026}, education~\citep{usta2021learning}, finance~\citep{choe-etal-2025-hierarchical}, and intellectual property~(IP)~\citep{roda-clefip-2009}.
With the rapid development of large language models~(LLMs), the paradigm of information access is gradually shifting from discriminative relevance estimation~\citep{TF-IDF,BM25} over in-domain corpora toward generative~\citep{li-survey-25} and agentic search paradigm~\citep{du2026openseeker,ning-agentic-26}. This shift is particularly consequential for patent examination, where reliable novelty assessment fundamentally depends on discovering and reasoning over evidence from large-scale prior-art corpora. 

Patent examination is inherently a knowledge-intensive information access task~\citep{wang2026patre}. 
To assess the novelty of an application, examiners must discover relevant evidence from massive collections of prior art rather than relying on pre-identified references.
They then analyze the retrieved prior art against claim limitations to determine whether the invention has been previously disclosed~\citep{Knappich-pat-2026}.
This goes beyond fine-grained relevance matching~\citep{risch2020patentmatch,ni-etal-2024-mozip} or reasoning~\citep{lim-PANORAMA,wang2026patre} over given prior-art documents: 
novelty assessment in patent examination requires discovering relevant prior art from a large search space and using intermediate findings to guide subsequent search, claim-level comparison, and generate an Office Action (OA)-style report under relevant legal constraints~\citep{SETCHI2021102021}.
Such a process naturally forms a tight \textbf{\textit{search--reasoning--assessment}} loop, making patent examination a particularly challenging setting for agentic search. 

Recent efforts have progressively shifted patent novelty assessment toward more realistic examination settings. 
HUPD~\citep{suzgun2023harvard} formulated patent examination as application-level acceptance prediction, while IPBench~\citep{wang-etal-2026-towards-ip} evaluated the patent examination and classification capabilities of LLMs. 
PANORAMA~\citep{lim-PANORAMA}, in contrast, decomposed claim-level examination into prior-art retrieval, relevant-paragraph identification, and novelty or non-obviousness classification.
More recent work further emphasizes fine-grained and process-oriented evaluation: FiNE-Patents~\citep{Knappich-pat-2026} assesses feature-level passage retrieval and novelty reasoning, while PatRe~\citep{wang2026patre} models the multi-turn lifecycle of \textit{OA}  generation and applicant rebuttal.
Existing benchmarks remain static and narrowly scoped relative to realistic examination: they evaluate examination components in isolation, constrain retrieval to small predefined candidate pools, or assume access to relevant prior art. As a result, they cannot capture the adaptive and iterative search, evidence discovery, and reasoning required in realistic patent examination.

 \begin{figure*}[!t]
      \centering
      \includegraphics[width=\textwidth]{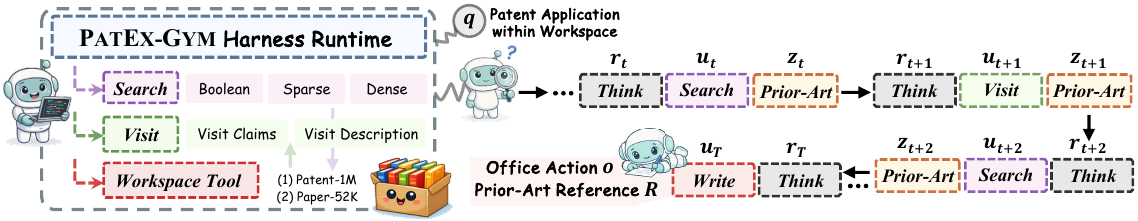}
    \caption{Overview of the Harness Runtime in PatEx-Gym, supporting iterative reasoning, prior-art retrieval, document inspection, and OA generation. $r_t$, $u_t$, and $z_t$ denote the reasoning, tool call, and tool results at step $t$, respectively.}   
    \label{fig:bench-framework}
  \end{figure*}

To address these gaps, we present \textbf{PatEx-Gym}, to the best of our knowledge, the first framework for developing and evaluating agents for novelty-oriented prior-art search and reasoning, with generated OAs serving as a downstream signal of novelty assessment.
Specifically, it integrates three core components that unify benchmarking, scalable trajectory synthesis, and supervised agent training within a single framework, as illustrated in Figure~\ref{fig:bench-framework}.
Beyond evaluating final OA alone, our PatEx-Gym exposes the complete \textit{search--reasoning--assessment} process, enabling both fine-grained search behavior analysis and the construction of high-quality trajectory for training stronger patent examination agents.
More importantly, PatEx-Gym also serves as a challenging testbed for general search agents, since its broad coverage across IPC sections naturally spans heterogeneous scientific and technical domains, requiring agents to perform cross-domain \textit{query adaptation}, \textit{tool use}, \textit{evidence integration}, \textit{technical reasoning}, and \textit{report generation}.
\textbf{Our main contributions are as follows:}
\begin{itemize}
\item We introduce \textbf{PatEx-Bench}, the first benchmark for agentic prior-art search and novelty assessment in patent examination scenario, comprising 1K realistic tasks. 
It's supported by large-scale \textit{Patent-1M}, \textit{Paper-52K}, and a diverse search toolset for prior-art search, reasoning, and OA generation.
\item We propose \textbf{PASS}, 
\textit{i.e.}, 
\textbf{\underline{P}}rior-\textbf{\underline{A}}rt Agentic \textbf{\underline{S}}earch Trajectory \textbf{\underline{S}}ynthesis, a annotation-free framework that synthesizes high-quality prior-art search trajectories. 
By treating document visits as implicit relevance signals and evidence-acquisition anchors, PASS models intra- and inter-turn search behavior and progression, together with rollout-level evidence and outcome quality, yielding \textbf{PatEx-SynData}, which contains 3K filtered trajectories for agent training.
\item Building on PatEx-SynData, we develop \textbf{PatEx-Agent} at \textbf{4B} scale via SFT for prior-art search and novelty assessment in patent examination. 
Despite its compact scale, our PatEx-Agent exhibits strong performance and establishes a lightweight baseline: the 4B variant performs comparably to models with \(\sim\)30B parameters and achieves Searched Recall~(Search-R) on par with Qwen3.7-Max.
This also demonstrates the potential of lightweight agents for efficient local deployment in practical patent examination settings.
\end{itemize}

\section{Related Work}

\subsection{Agentic Search and Deep Research}

Conventional information retrieval methods estimate the relevance between a user query and documents in a corpus, returning a ranked list through lexical or learned matching functions~\citep{BM25,burges-l2r-05}. 
With the emergence of LLMs, search systems increasingly move beyond scoring for a fixed query toward actively planning and adapting the search process~\citep{li-survey-25}. 
ReAct~\citep{yao2023react} and IRCoT~\citep{trivedi-etal-2023-interleaving} provide foundations for interleaving reasoning with environment actions or retrieval. 
Recent work further focuses on training search agents.
Search-R1~\citep{jin2025searchr} learns multi-turn reasoning and search through reinforcement learning and TIPS~\citep{xie2026tips} leverage turn-level reward to optimize search-augmented LLMs.
FORT-Searcher~\citep{deng2026fort} constructs shortcut-resistant search tasks and trajectories. 
And SearchArt~\citep{mei2026searchart} combines verified task synthesis with multi-stage post-training.
These methods formulate search as a sequential decision process in which retrieved evidence informs subsequent queries and reasoning.

\begin{wraptable}{r}{0.52\linewidth}
\vspace{-1em}
\centering
\caption{Comparison of patent examination benchmarks. Evidence Pool denotes the searchable candidates. Where PatEx-Bench comprises 1M patents and 52K papers. \textit{Reason} denotes LLMs reasoning beyond retrieval alone; \textit{OA} denotes Office Action generation; \textit{Agentic} indicates whether provides a harness for iterative tool use and multi-step task execution, and \textit{NPL} denotes support for non-patent literature.}
\label{tab:patent-benchmark-comparison}

\setlength{\tabcolsep}{1.5pt}
\renewcommand{\arraystretch}{1.05}

\resizebox{\linewidth}{!}{%
\begin{tabular}{@{}l l l c c c c@{}}
\toprule
\textbf{Benchmark} &
\textbf{Test~\#} &
\textbf{Evidence Pool} &
\textbf{Reason} &
\textbf{OA} &
\textbf{Agentic} &
\textbf{NPL} \\
\midrule
PAR4PC
& 2,896
& 8 candidates
& \ding{51}
& \ding{55}
& \ding{55}
& \ding{55} \\

IPBench Task~3--5
& 314
& no prior art
& \ding{51}
& \ding{55}
& \ding{55}
& \ding{55} \\

FiNE-Patents
& 1,844
& 1 given document
& \ding{51}
& \ding{55}
& \ding{55}
& \ding{55} \\

PatRe
& 480
& 10 references
& \ding{51}
& \ding{51}
& \ding{55}
& \ding{55} \\
\midrule

\textbf{PatEx-Bench~\textit{(Ours)}}
& 1,000
& 1M + 52K
& \ding{51}
& \ding{51}
& \ding{51}
& \ding{51} \\
\bottomrule
\end{tabular}%
}
\vspace{-0.5em}
\end{wraptable}
Agentic search has further expanded to long-horizon Web research~\citep{wei2025browsecomp} and specialized document retrieval. 
Deep Research~(DR)~\citep{shi2025deep} represents a prominent form of agentic search, in which agents conduct multi-step retrieval, synthesize evidence across sources, and produce citation-grounded answers or reports.
Recent DR benchmarks and methods evaluate and train agents in terms of answer quality, citation correctness, and search behavior~\citep{li2026openresearcher,team2025tongyi,ICLR2026_465f22be,gou2026mindweb}.
In specialized settings, PaperScout~\citep{pan2026paperscout} dynamically combines direct search with citation expansion for academic literature retrieval.
Patent examination requires search to support a formal, claim-level decision under strict evidentiary and legal constraints.
Our \textbf{PatEx-Gym} adapts the agentic search paradigm to patent examination, where agents search a large-scale prior-art corpus and connect iterative evidence discovery with claim-level reasoning to produce traceable examination. 

\subsection{Patent Retrieval and Examination}
Prior-art retrieval is fundamental to patent examination, as reliable novelty assessments depend on identifying relevant earlier disclosures from massive patent collections. 
NTCIR Patent Retrieval~\citep{fujii-etal-2004-test} and CLEF-IP~\citep{clefip11} establish benchmarks for prior-art search, catalyzing advances in lexical retrieval, query expansion, cross-lingual matching, and patent-specific ranking. 
Within the broader examination process, retrieval is interleaved with query refinement, inspection of full disclosures, and reasoning that connects prior-art evidence to legally grounded decisions.

Beyond ranking-based evaluation of prior-art retrievers~\citep{fujii-etal-2004-test,suzgun2023harvard}, recent benchmarks incorporate finer-grained evidence and reasoning into patent examination using LLMs~\citep{wang-etal-2026-towards-ip}. 
PANORAMA~\citep{lim-PANORAMA} decomposes claim-level examination into prior-art retrieval, paragraph identification, and novelty classification. 
However, these stages are evaluated separately and retrieval is restricted to a small and fixed candidate pool.
FiNE-Patents~\citep{Knappich-pat-2026} provides feature-level annotations for passage retrieval and novelty reasoning, but limited in a given prior-art document.
PatRe~\citep{wang2026patre} models OA generation and applicant rebuttal under various reference settings. 
As summarized in Table~\ref{tab:patent-benchmark-comparison}, \textbf{PatEx-Bench} advances toward an agentic paradigm by evaluating the complete \textit{search--reasoning--assessment} trajectory, where agents autonomously search across large-scale prior-art corpora.

\section{PatEx-Bench}
\label{sec:patex-bench}

In this section, we introduce \textbf{PatEx-Bench}, the first benchmark for evaluating agentic prior-art search and novelty assessment in patent examination scenario.
Unlike settings in fixed candidate documents, PatEx-Bench requires an agent to autonomously discover temporally eligible patents and non-patent literature~(NPL) and inspect their underlying disclosures. 
The agent must then reason over claims and produce an evidence-grounded novelty assessment in the form of a traceable Office Action~(OA).
Each task runs in an isolated Codex harness that provides the workspace and toolsets over our large-scale Patent-1M and Paper-52K corpora. 
This design makes it possible to evaluate not only the final OA, but also the complete \textit{search--reasoning--assessment} trajectory.

\subsection{Problem Definition}
\label{sec:patex-task}

Let $\mathcal{Q}=\{q_i\}_{i=1}^{N}$ denote a set of patent applications, and let $\mathcal{C}=\mathcal{P}\cup\mathcal{N}$ denote a large-scale prior-art corpus consisting of patent documents $\mathcal{P}$ and NPL $\mathcal{N}$ (scientific papers). 
Each $q_i$ represents a patent application and includes its bibliographic metadata, title, abstract, specification, claims, and filing date $d_i$.
To enforce the legal temporal constraint on prior-art eligibility, we restrict the searchable corpus for $q_i$ to documents published before $d_i$, defining $\mathcal{C}_i=\{e\in\mathcal{C}\mid d_e<d_i\}$, where $d_e$ denotes the date of document $e$.

Given $(q_i,\mathcal{C}_i)$, an agent $\pi$ follows a ReAct-style interaction pattern~\citep{yao2023react}, repeatedly alternating between reasoning and calls to a set of tools provided by the harness to produce a multi-turn agentic search trajectory $\tau_i$, i.e.,
\begin{equation}
    \tau_i=\bigl\{(r_{i,t},u_{i,t};z_{i,t})\bigr\}_{t=1}^{T_i},
    \qquad ({o}_i,{R}_i)=\pi(q_i,\mathcal{C}_i;\tau_i).
\end{equation}
At each step $t$, the reasoning state $r_{i,t}$ guides a tool call $u_{i,t}$, while $z_{i,t}$ denotes the tool result returned by the harness. 
The agent decides whether to continue invoking tools or terminate the trajectory by submitting an OA ${o}_i$ and its reference set ${R}_i\subseteq\mathcal{C}_i$. 

\subsection{Auditable Execution Harness Runtime}
\label{harness-runtime}
As shown in Figure~\ref{fig:bench-framework}, we construct a sandbox-based execution environment using harbor~\citep{Harbor_Framework}, equipping them with a diverse prior-art search toolset, including Boolean keyword search, sparse retrieval~(BM25~\citep{BM25}), and dense retrieval~(BGE-M3~\citep{chen2024m3}), together with basic workspace tools, e.g., \textit{Read} and \textit{Write}. 
Notably, the search tools return only the bibliographic metadata, titles, and abstracts of documents.
We further provide visit tools to support progressive disclosure of prior-art information, allowing agents to dynamically inspect documents they deem more relevant or potentially novelty-defeating and access richer content such as claims. 
Therefore, the agent primarily operates through 3 action types, \textbf{\textit{Reasoning}}, \textbf{\textit{Search}}, and \textbf{\textit{Visit}}.

Based on these tool designs, we integrate the search capabilities with the \leaderlogo{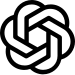}~Codex~\citep{codex} harness, enabling PatEx-Gym to support robust agentic runtimes.
To ensure fair and reliable evaluation, we divide each agentic runtime into three strictly separated stages: Preparation, Execution, and Evaluation. 
During task execution, the agent has no access to the evaluation rubrics. 
Finally, the harness records the complete and auditable agent trajectory, search-tool execution logs, final environment state, and the requested artifacts.

\subsection{Evaluation Criteria}
\label{metrics}
Evaluating agentic patent examination is challenging because success depends not only on the quality of the final generated OA, but also on the correctness of its legal conclusions and the agent's prior-art search process. 
We introduce a multi-dimensional evaluation protocol covering document quality, legal decision accuracy, and search-to-select funnel-style prior-art recall. 

\textit{(1) LLM-judge-based Office Action Quality.}
We use a format score $F_i\in[0,1]$ to assess the file presence and structural validity of the artifacts.
Three distinct LLM judges independently evaluate the generated OA $o_i$ against the ground-truth OA $\hat{o}_i$, assigning scores from 1 to 10 along five dimensions, following~\citet{wang2026patre}: \emph{soundness}, \emph{clarity}, \emph{constructiveness}, \emph{completeness}, and \emph{language style}.
We average the scores across judges and dimensions and normalize them to obtain an OA quality score $Q_i(o_i,\hat o_i)\in[0,1]$. The overall quality score of generated OA is
\begin{equation}
\operatorname{Score}=\mathbb{E}_{q_i\sim\mathcal{Q}}\left[\frac{1}{6}F_i+\frac{5}{6}Q_i(o_i,\hat o_i)\right]\times 10.
\end{equation}

\textit{(2) Decision and Statutory-ground Accuracy.}
To assess substantive correctness beyond document quality, we measure the agreement between the generated OA and the ground-truth prosecution outcome at two levels: the final decision and the applicable statutory grounds. 
We quantify these aspects using Decision Accuracy~(Dec.) and Statute F1~(Stat.), respectively.
For each generated OA $o_i$, we extract its predicted decision $y_i$ and statutory-ground set $B_i$, and compare them with the ground-truth labels $\hat{y}_i$ and $\hat{B}_i$:
\begin{equation}
 \begin{aligned}
      \mathrm{Dec.}&= \mathbb{E}_{q_i\sim\mathcal{Q}}
      \bigl[\mathbb{I}[y_i=\hat{y}_i]\bigr],& 
      \mathrm{Stat.}&= \mathbb{E}_{q_i\sim\mathcal{Q}}
      \left[\frac{2|B_i\cap\hat{B}_i|}
      {|B_i|+|\hat{B}_i|}\right].
  \end{aligned}
\end{equation}
Where $\mathbb{I}[\cdot]$ denotes the indicator function.

\textit{(3) Prior-art Discovery and Select.}
We trace the examiner-cited prior art through successive stages of the agent's search trajectory, from retrieval and inspection to final selection in the OA. 
For both patents and NPL, let ${S}_i$ denote the documents returned by search-tool calls using queries formulated by the agent, ${V}_i$ the documents accessed through the visit tool, and ${R}_i$ the references substantively selected in the generated OA.
Using the ground-truth reference sets $\hat{R}_i$, we evaluate this retrieval-to-use funnel-style recall as
\begin{equation}
\begin{gathered}
    \operatorname{Search\text{-}R} = \mathbb{E}_{q_i\sim\mathcal{Q}}\left[\frac{|\hat{R}_i\cap{S}_i|}{|\hat{R}_i|}\right],
    \qquad
    \operatorname{Visit\text{-}R} = \mathbb{E}_{q_i\sim\mathcal{Q}}\left[\frac{|\hat{R}_i\cap{V}_i|}{|\hat{R}_i|}\right], \\
    \operatorname{Select\text{-}R} = \mathbb{E}_{q_i\sim\mathcal{Q}}\left[\frac{|\hat{R}_i\cap{R}_i|}{|\hat{R}_i|}\right].
\end{gathered}
\end{equation}
where the Searched Recall~(Search-R), Visited Recall~(Visit-R), and Selected Recall~(Select-R) measure the recall of examiner-cited references $\hat R_i$ at the retrieval, access, and final-use stages, respectively. 

\subsection{Data Collection, Processing and Statistics}

\paragraph{Data Collection and Processing.}
PatEx-Bench is grounded in verifiable patent prosecution records sourced from publicly available USPTO research datasets
, which contains examiner-authored OAs. 
As official examination decisions issued by qualified USPTO examiners, these OAs carry procedural legal effect and provide authoritative, professionally validated golden labels rather than synthetic annotations.
For each $q_i$, we retain the first OA and the application in its originally filed version, including its bibliographic metadata, abstract, specification, and claims, ultimately yielding \textbf{{1K}} tasks.
This preserves the initial claims available to the agent instead of the final granted document.
To support prior-art search, we further construct \textbf{\textit{Patent-1M}} and \textbf{\textit{Paper-52K}} from publicly accessible Google Patent and Scholar, respectively.
We standardize heterogeneous identifiers and fields into a unified schema and remove duplicate records. 
Notably, we verify that every examiner-cited oracle reference for each OA is retrievable from these two corpora.

\begin{figure*}[!t]
    \centering
    \begin{minipage}[t]{0.245\textwidth}
        \centering
    \includegraphics[width=\linewidth]{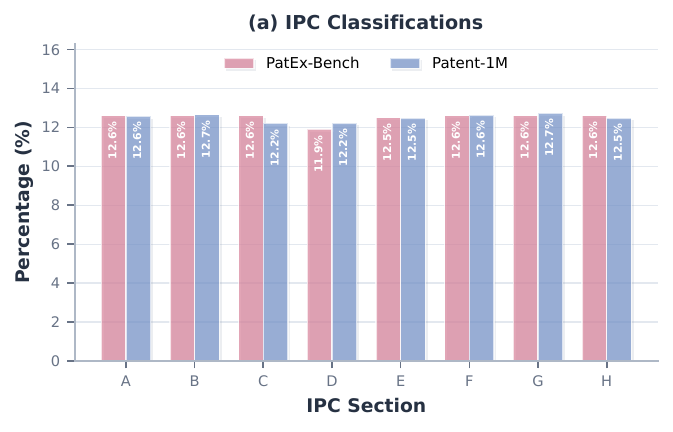}
    \end{minipage}\hfill
    \begin{minipage}[t]{0.245\textwidth}
        \centering
    \includegraphics[width=\linewidth]{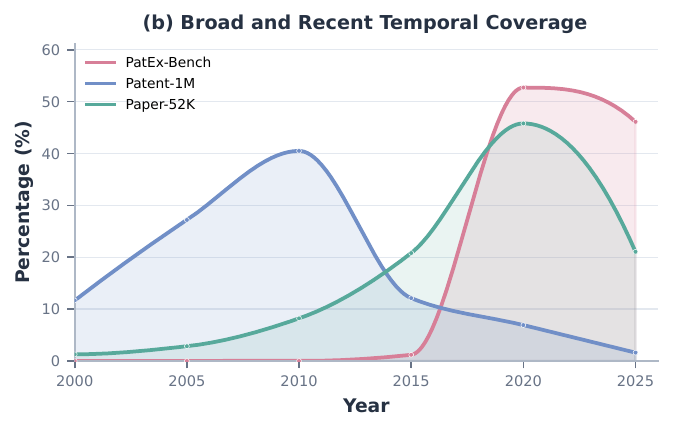}
    \end{minipage}\hfill
    \begin{minipage}[t]{0.245\textwidth}
        \centering
    \includegraphics[width=\linewidth]{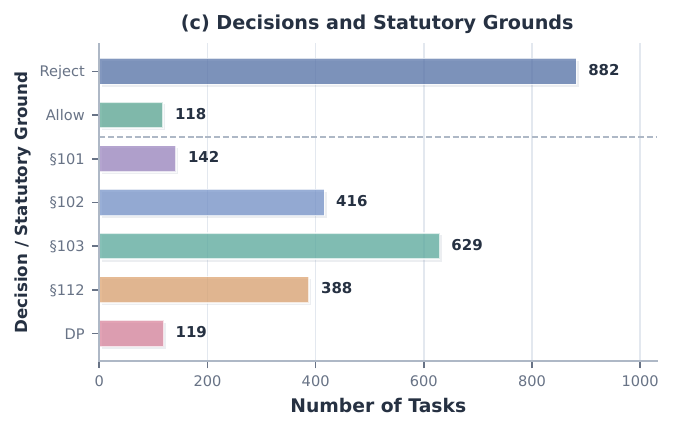}
    \end{minipage}\hfill
    \begin{minipage}[t]{0.245\textwidth}
        \centering
    \includegraphics[width=\linewidth]{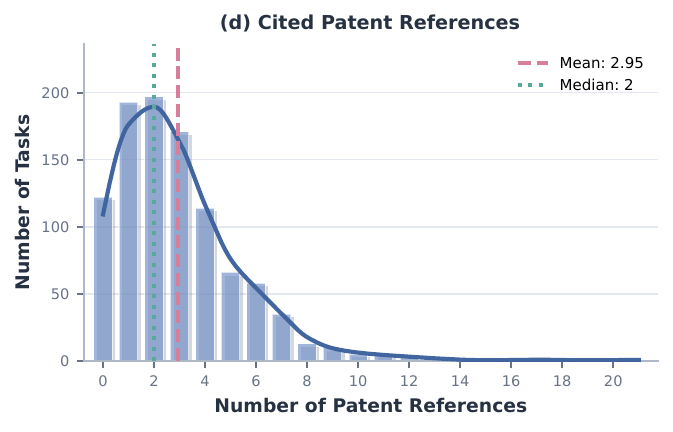}
    \end{minipage}
    \caption{Data distributions for PatEx-Bench and its retrieval corpora. International Patent Classification~(IPC) including: A, human necessities; B, performing operations and transporting; C, chemistry and metallurgy; D, textiles and paper; E, fixed constructions; F, mechanical engineering, lighting, heating, weapons, and blasting; G, physics; and H, electricity.}
    \label{fig:patex-data-statistics}
\end{figure*}

\paragraph{Data Statistics.}
Our PatEx-Bench comprises 1,000 patent examination tasks spanning all eight IPC sections, with 2,946 examiner-cited patents and 178 examiner-cited papers serving as ground-truth prior-art references.
The \textit{Patent-1M} and \textit{Paper-52K} corpora contain 1,011,901 unique patent documents and 52,105 scientific papers, respectively.
This poses a substantial challenge for agents to efficiently discover and verify relevant prior art.
As shown in Figure~\ref{fig:patex-data-statistics}, we present the distributions of (a) IPC classifications, (b) temporal coverage, (c) decisions and statutory grounds, and (d) examiner-cited patent references for PatEx-Bench. 

 \begin{figure*}[!t]
      \centering
      \includegraphics[width=\textwidth]{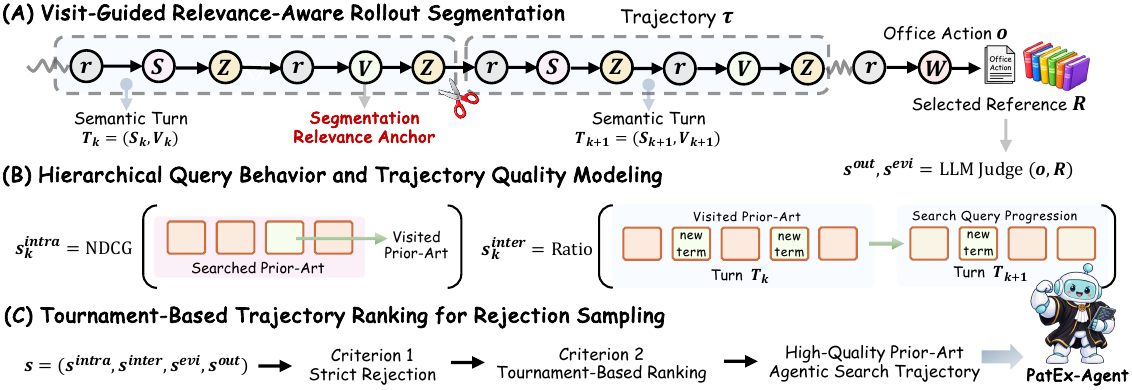}
    \caption{Overview of Prior-Art Agentic Search Trajectory Synthesis~(PASS) for PatEx-SynData. 
    (A) Visit-guided, relevance-aware rollout segmentation divides trajectories into semantic turns using visits as anchors. $r$ denotes reasoning, $S$ denotes search, $V$ denotes visit, $Z$ denotes the returned tool results, and $W$ denotes workspace tools, e.g. \texttt{Write}.~(B) Hierarchical quality modeling evaluates multidimensional quality.
    (C) Tournament-based ranking with rejection sampling for high-quality trajectories.}   
    \label{fig:PASS-framework}
  \end{figure*}

\section{PatEx-SynData and PatEx-Agent}

Building upon the workspace environment of PatEx-Gym, we further investigate whether strong prior-art search capabilities can be acquired through agentic training. 
We propose \textbf{PASS}, a scalable trajectory synthesis framework that generates prior-art search trajectories with a teacher model and leverages document visits as implicit relevance signals to segment trajectories, model fine-grained trajectory quality, and curate high-quality search behaviors.
We further use PASS to construct \textbf{PatEx-SynData}, a curated training set of approximately 3K high-quality prior-art search trajectories selected from 10K teacher-generated rollouts. 
Based on PatEx-SynData, we train \textbf{PatEx-Agent} to acquire effective agentic search behaviors for prior-art discovery in realistic patent examination scenario.

\subsection{Prior-Art Search Trajectory Synthesis}

In this section, we introduce \textbf{PASS}, \textit{i.e.}, \textbf{\underline{P}}rior-\textbf{\underline{A}}rt Agentic \textbf{\underline{S}}earch Trajectory \textbf{\underline{S}}ynthesis, a scalable, annotation-free framework that synthesizes prior-art search trajectories with a teacher model using only structured patent documents as inputs, and curates them based on fine-grained quality at both the turn and trajectory levels.
Specifically, as shown in Figure~\ref{fig:PASS-framework}, PASS leverages the progressive disclosure of patent information through visit tools to enable visit-aware, turn-level trajectory quality ranking. 
It treats document visits as implicit relevance signals and evidence-acquisition anchors to capture both retrieval-to-inspection efficiency and evidence-driven search progression for novelty assessment in patent examination. 

\subsubsection{Visit-Guided Relevance-Aware Rollout Segmentation} 
Given an application $q_i$, we employ a teacher model to interact with the harness and generate a complete prior-art search and OA generation trajectory $\tau_i$. 
This synthesis process is \textit{annotation-free} and requires no examiner-cited references, enabling scalable trajectory generation directly from large-scale structured patent collections. 
In the progressive search-then-visit paradigm of PatEx-Gym, each visit not only reflects an implicit relevance judgment over the retrieved candidates~\citep{LRAT}, analogous to clickthrough-based implicit feedback in Web search~\citep{click-Joachims}, but also marks the acquisition of richer document-level evidence.
Based on this observation, PASS views visit actions as natural anchors and performs visit-guided relevance-aware segmentation to decompose each raw trajectory $\tau_i$ into semantically coherent search turns, \textit{i.e.},
\begin{equation}
\tau_i=\{T_1,T_2,\ldots,T_K\}, \qquad
T_k=\bigl(\mathcal{S}_k,\mathcal{V}_k\bigr).
\end{equation}
where the $k$-th search turn $T_k$ as a sequence of search actions followed by visit actions. 
Notably, the agent may invoke multiple search tools or visit tools in parallel within a single tool call. 
Therefore, we denote the sequences of search and visit actions in the $k$-th turn as $\mathcal{S}_k=\{S_{k,1},\ldots,S_{k,m_k}\}$ and $\mathcal{V}_k=\{V_{k,1},\ldots,V_{k,n_k}\}$ respectively. 
A new turn begins when the agent issues a new search after consuming the visited evidence from the preceding turn. 

\subsubsection{Hierarchical Query Behavior and Trajectory Quality Modeling}
After the teacher model rollout and produces a complete trajectory $\tau_i$, it may exhibit substantial variation in search efficiency, query adaptation, evidence grounding, and final examination quality.
Simply retaining trajectories based on successful outcomes therefore cannot distinguish whether the agent reaches its conclusion through effective and reliable search behavior. 
To capture fine-grained attribution, our PASS performs hierarchical quality modeling over the complete rollout, jointly characterizing intra-turn query behavior, inter-turn search progression, and the quality of evidence and outcome, \textit{i.e.,} for $T_k$ and  $T_{k+1}$ turns as
\begin{equation}
\begin{aligned}
\text{Turn-Level}\quad&
s_i^{\mathrm{intra}}: \overbrace{\mathcal{S}_k \rightarrow \mathcal{V}_k}
^{\substack{\text{Intra-turn}\\\text{Query Behavior}}}
,
&s_i^{\mathrm{inter}}: \overbrace{\mathcal{S}_k \rightarrow \mathcal{V}_{k} \rightarrow \mathcal{S}_{k+1}}^{\substack{\text{Inter-turn}\\\text{Search Progression}}}.\\
\text{Trajectory-Level\quad}&s_i^{\mathrm{evi}}: \underbrace{\mathcal{V}_{1:K} \rightarrow {R}_i}_{\substack{\text{Evidence Grounding}}}
,
&s_i^{\mathrm{out}}: \underbrace{(R_i,q_i) \rightarrow y_i}_{\substack{\text{Outcome Quality}}}.
\end{aligned}
\end{equation}
where $s_i^{\mathrm{intra}}$ and $s_i^{\mathrm{inter}}$ represent search behavior within and across turns, respectively, $s_i^{\mathrm{evi}}$ and $s_i^{\mathrm{out}}$ evaluate evidence grounding and final examination quality at the trajectory level.

\paragraph{Turn-Level Query Behavior and Search Progression Quality.}
We analyze the concrete search behavior in a trajectory through intra- and inter-turn analysis. 
\textbf{(1)} At each turn $T_k$, the \textit{intra-turn} score $s_{i,k}^{\mathrm{intra}}$ measures how efficiently the agent turns retrieved candidates into inspections, computed via NDCG over visited documents against the search results.
\textbf{(2)} For turn pair $(T_k, T_{k+1})$, the \textit{inter-turn} score $s_{i,k}^{\mathrm{inter}}$ captures how the search query meaningfully evolves across turns, measuring whether $\mathcal{S}_{k+1}$ reflects a substantive shift or refinement from $\mathcal{S}_k$ in response to the information acquired in $\mathcal{V}k$, rather than merely rephrasing the previous query. 
A high $s{i,k}^{\mathrm{inter}}$ indicates clear query progression, such as introducing new concepts based on newly acquired evidence.

\paragraph{Trajectory-Level Evidence Grounding and Outcome Quality.}
Beyond search behavior, we further assess the quality of the final OA and the evidence grounding of the selected references.
\textbf{(1)} We leverage the \textit{evidence} score $s_i^{\mathrm{evi}}$ measures that every entry in the selected reference set $\mathcal{E}_i$ is grounded in the content the agent actually inspected. 
Specifically, we leverage an LLM to tell whether the office-action statements referring to it are supported by its visited passages. 
\textbf{(2)} The \textit{outcome} score $s_i^{\mathrm{out}}$ evaluates the final OA $y_i$ without ground-truth reference. 
It leverages an LLM to evaluate two aspects: whether each claim is explicitly resolved with appropriate justification, and whether the references supporting statutory rejections are properly grounded in the selected evidence.

\subsubsection{Tournament-Based Trajectory Ranking for Rejection Sampling}
\label{ranking-traj}
Given a trajectory pool $\mathcal{T}=\{\tau_i\}_{i=1}^{N}$, where each trajectory $\tau_i$ is scored with a hierarchical quality $\mathbf{s}_i=(s_i^{\mathrm{intra}}, s_i^{\mathrm{inter}}, s_i^{\mathrm{evi}}, s_i^{\mathrm{out}})$, PASS selects high-quality and trainable trajectories from $\mathcal{T}$ via two stages: first, a strict rejection gate that discards clearly unusable rollouts~(\textit{e.g.}, missing required deliverables or failing to execute the task), followed by a tournament-based ranking over score. Detailed trajectory filtering and ranking criterion are as follows:

\textbf{(I)~Criterion 1: Strict Rejection.} We apply a deterministic feasibility gate that requires each trajectory to provide both the Office Action $y_i$ and selected references $R_i$, and remain within the allowed turn budget. Trajectories failing any condition are discarded as incomplete, non-agentic, pathological, or judge-failed rollouts.

\textbf{(II)~Criterion 2: Tournament-Based Ranking.} Trajectories surviving strict rejection are ranked by a pairwise tournament over $\mathbf{s}_i$. 
    For each pair $(\tau_i,\tau_j)$, we compare their four-dimensional score and declare $\tau_i$ the winner if it outperforms $\tau_j$ on a strict majority of four dimensions:
    \begin{equation}
    \tau_i \succ \tau_j
    \iff
    \sum_{d\in\mathcal{D}}
    \mathbb{I}[s_i^d>s_j^d]>
    \frac{|\mathcal{D}|}{2}.
    \end{equation}
    where $\mathcal{D}=\{\mathrm{intra},\mathrm{inter},\mathrm{evi},\mathrm{out}\}$. The tournament score $W_i$ for ranking is the total number of pairwise wins,
    \begin{equation}
    W_i=\sum_{j\neq i}\mathbb{I}[\tau_i\succ\tau_j],
    \end{equation}
This aggregation preserves the vector structure of $\mathbf{s}_i$ without collapsing it into a scalar, keeping distinct failure modes separable and enabling ranking-based rejection sampling for trajectory selection.

\subsection{Agent Training}

\subsubsection{Data Recipes}

\begin{wrapfigure}{r}{0.55\linewidth}
    \centering
    \vspace{-1em}
    \includegraphics[width=\linewidth]{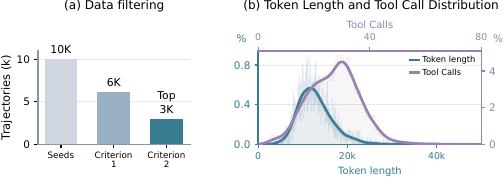}
    \caption{Training data filtering and the distributions of token length and tool calls.}
    \vspace{-1em}
    \label{fig:data-curation}
\end{wrapfigure}
Given the annotation-free nature of PASS, we crawl \textit{\textbf{10K seed patents}} from the USPTO, balanced across all IPC sections and strictly disjoint from PatEx-Bench.
Using the Codex harness of PatEx-Gym with the predefined search and visit tools, we synthesize trajectories by allowing the teacher model to reason, invoke tools, and generate outputs within the harness workspace, resulting in 10K training trajectories.
Crucially, the teacher model has no access to the trajectory-quality evaluation criteria during generation. 
Then, we apply PASS to segment the rollouts and compute their turn- and trajectory-level quality scores after filtering out invalid trajectories using Criterion~1. 
After filtering, we retain 6,321 trajectories within the admissible budget and rank them using the tournament-based procedure described in Section~\ref{ranking-traj}. 
Finally, we select the top $\sim$3K high-quality trajectories according to their tournament scores as the training data, termed \textbf{PatEx-SynData}. 
As shown in Fig.~\ref{fig:data-curation}, we provide an overview of the PASS data curation process: (a) illustrates the progressive filtering of teacher-generated trajectories from the initial 10K seeds through the two selection criteria, while (b) presents the token-length and tool-call distributions of the resulting trajectories.

\subsubsection{Training Recipes}
For a training trajectory from our PatEx-SynData  $\tau=(a_1,z_1,a_2,z_2,\ldots,a_T,z_T)$, where $a_t$ denote the model output (reasoning $r_t$ and tool calls $u_t$), and $z_t$ refers to harness-returned tool results at turn $t$, respectively.
We perform supervised fine-tuning~(SFT) on the model with environment masking, \textit{i.e.}
\begin{equation}
\mathcal{L}_{\mathrm{SFT}}
=
-\sum_{t=1}^{T}
\sum_{j=1}^{|a_t|}
\log p_{\theta}
\left(
a_t^{(j)}
\mid
\,a_t^{(<j)},\tau_{<t}
\right)
\label{loss}
\end{equation}
where the loss is computed only over output tokens $a_t^{(j)}$, conditioned on the preceding context $(a_t^{(<j)},\tau_{<t})$, while all system, user, and harness-message tokens are masked from the training objective.

\section{Experiment}

\subsection{Experimental Setup}
\label{experimental-setup}

\paragraph{Baselines.}
We evaluate 13 models spanning different families and parameter sizes, including \textbf{(1)} Frontier models: Claude-Opus-5, Claude-Opus-4.7, GLM-5.3, DeepSeek-V4-Flash-0731~\citep{xu2026deepseek}, Qwen3.7-Max, Qwen3.8-Max-0813~\citep{qwen37,qwen38} and Kimi-K2.6; \textbf{(2)} Open-source models spanning both dense and MoE architectures across various parameter scales, including Qwen3.5-4/9/397A17B and Qwen3.6-27/35A3B. 
We also evaluate the retrieval performance of 12 retrievers and the agents on the patent corpus of our PatEx-Bench, including \textbf{(1)} Sparse Retriever: BM25~\citep{BM25}, TF-IDF~\citep{TF-IDF}, uniCOIL~\citep{lin2021few} and SPLADE-v3~\citep{lassance2024splade}; \textbf{(2)} Dense Retriever: BGE-M3~\citep{chen2024m3}, jina-embedding-v5-text-nano/small~\citep{akram2026jina}, Qwen3-Embedding-0.6/4/8B~\citep{zhang2025qwen3} and domain-specific models including PaECTER~\citep{ghosh2024paecter} and PatEmbed-Large~\citep{ayaou2025patenteb}.

\paragraph{Implementations.} For agent training, we optimize the SFT objective in Equation~\ref{loss} using Qwen3.5-4B as the backbone and the PatEx-SynData synthesized by PASS with GLM-5.3 as the teacher model. 
We use the slime~\citep{slime_github} framework with a reproducible configuration and train the model for 3 epochs on 8 NVIDIA H20 GPUs with a learning rate of $1\times10^{-5}$. 
For fair evaluation on PatEx-Bench, we use official APIs for closed-source models and larger open-source models, while deploying smaller models, such as Qwen3.5-9B, locally with the vLLM~\citep{kwon2023efficient}.
All models are evaluated under the Codex~(v0.151.0) with a temperature of 1.0, a maximum output length of 32,768 tokens, and a maximum of 50 steps. The harness retains its native context-compaction mechanism. For models that support configurable reasoning effort, we use the highest available setting (e.g., \textit{xhigh} or \textit{max}), indicated by the corresponding postfix.
We use the evaluation criteria defined in Section~\ref{metrics} to provide a multi-dimensional assessment of prior-art search performance, averaging scores from 3 LLM judges, \textit{i.e.}, \textit{Claude-Opus-4.7}, \textit{Gemini-3.5-Flash}, and \textit{GPT-5.5}. All judges use a temperature of 0 and a maximum output length of 8,192 tokens.

\subsection{Main Results on PatEx-Bench}

As shown in Table~\ref{tab:main_results}, we report the performance of evaluated models on PatEx-Bench under the Codex harness.
Overall, the agents exhibit substantial performance variation, with Scores ranging from 3.79 to 7.34.
Claude-Opus-5-max achieves the best results in Score, Statute F1, and all three reference-recall metrics.
The Select-R peaks at only 32.4, highlighting substantial room for improvement in statutory reasoning and prior-art search.
Notably, despite its compact 4B scale, PatEx-Agent achieves a Search-R of 38.7, surpassing larger models such as Qwen3.5-397B-A17B, while remaining competitive in OA quality.
We further analyze the differences in terms of prior-art discovery and examination quality as follows.

\paragraph{Strong models still face substantial prior-art discovery challenges.}
Claude-Opus-5-max achieves the highest OA quality Score at 7.34, together with the strongest statute identification and reference recall across all three retrieval stages.
Nevertheless, even the strongest Search-R remains below 50\% and Select-R below 33\%, highlight the difficulty of discovering and retaining examiner-cited references through a progressive search process.
Notably, our PatEx-Agent substantially improves over its Qwen3.5-4B backbone, increasing the OA quality Score from 3.79 to 5.41~(\upperformence{42.7\%}) and yielding consistent gains across all search and reasoning metrics, ranging from \upperformence{17.5\%} to \upperformence{91.7\%}.
These results suggest that effective prior-art discovery is not solely determined by model scale, but can be substantially improved by explicitly training search and reasoning capabilities.

\begin{table}[t]
  \centering
  \caption{
    Main results on PatEx-Bench. Score is on a 0--10 scale. Decision Accuracy~(Dec.) and Statute F1~(Stat.) evaluate the final examination decision and statutory grounds, respectively. Searched~(Search-R), Visited~(Visit-R), and Selected~(Select-R) report the recall~(\%) at the corresponding stages. Best results in each column are \textbf{bolded}; second-best results are \underline{underlined}.
  }
  \label{tab:main_results}
  \setlength{\tabcolsep}{10pt}
  \resizebox{\columnwidth}{!}{
    \begin{tabular}{lcccccc}
      \toprule
      \textbf{Model}
      & \textbf{Score}
      & \textbf{Dec.}
      & \textbf{Stat.}
      & \textbf{Search-R}
      & \textbf{Visit-R}
      & \textbf{Select-R} \\
      \midrule
      \multicolumn{7}{
  >{\columncolor{gray!20}}c
}{
  \textbf{\leaderlogo{Figures/openai_logo.png}~Codex Harness}
} \\

      Claude-Opus-5-max
      & \textbf{7.34} & \underline{87.0} & \textbf{56.9} & \textbf{48.7} & \textbf{35.6} & \textbf{32.4}\\

      Qwen3.8-Max-xhigh
      & \underline{6.76} & 85.6 & {54.8} & 41.4 & \underline{30.1} & \underline{29.1} \\

      GLM-5.3-max
      & 6.71 & 86.7 & 54.5 & 40.4 & 29.4 & 28.4  \\

      Claude-Opus-4.7-max
      & 6.55 & 86.6 & \underline{55.6} & 41.6 & 28.7 & 26.2  \\

      DeepSeek-V4-Flash-0731-max
      & 6.17 & 79.9 & 50.2 & \underline{44.1} & 26.8 & 24.7  \\

      Qwen3.7-Max-xhigh
      & 5.98 & 84.1 & 50.0 & 36.1 & 25.9 & 24.1  \\

      Qwen3.6-27B
      & 5.96 & \textbf{87.6} & 51.6 & 35.5 & 21.8 & 19.5  \\

      Qwen3.5-397A17B
      & 5.55 & 83.7 & 47.5 & 30.4 & 17.3 & 16.8  \\

      Qwen3.6-35A3B
      & 5.55 & 84.9 & 45.7 & 35.3 & 20.6 & 20.3  \\

      Kimi-K2.6-max
      & 5.27 & 76.4 & 43.8 & 37.8 & 24.7 & 19.3  \\

      Qwen3.5-9B
      & 4.55 & 81.9 & 46.4 & 27.6 & 15.5 & 14.4  \\

      Qwen3.5-4B
      & 3.79 & 69.7 & 38.9 & 23.2 & 12.1 & 12.0  \\

      \multicolumn{7}{
  >{\columncolor{gray!10}}c
}{
  \textbf{\textit{Ours}}
} \\
      PatEx-Agent-4B
      & 5.41 & 82.9 & 45.7 & 38.7 & 23.2 & 18.2  \\
      \qquad\text{ }$\mathbf{\Delta}$ vs. Base & \upperformence{42.7\%} & \upperformence{18.9\%} & \upperformence{17.5\%} & \upperformence{66.8\%} & \upperformence{91.7\%} & \upperformence{51.7\%}  \\
      \bottomrule
    \end{tabular}
  }
\end{table}

\paragraph{Decision accuracy alone does not capture examination quality.}
Qwen3.6-27B achieves the highest Decision Accuracy of 87.6\%, yet its Score and Select-R remain substantially below those of Claude-Opus-5-max, at 5.96 vs. 7.34 and 19.5 vs. 32.4, respectively. 
Decision Accuracy also shows only moderate rank agreement with the Score, with Kendall's $\tau$ of 0.550. 
These results indicate that correct final decisions do not necessarily imply equally strong prior-art discovery or overall OA quality.
Additionally, we compute the agreement among the three LLM judges, obtaining Kendall's $W=0.849$, which demonstrates strong consistency in their quality rankings and supports the reliability of our LLM-based evaluation.

\begin{table}[!t]
\centering
\caption{Retrieval results on the Patent Corpus. Sparse and dense retrievers are evaluated using R@$K$. For each agent on \leaderlogo{Figures/openai_logo.png} Codex Harness~(\textit{w/} our standard toolset, \textit{i.e.}, Boolean, BM25 and BGE-M3), we order the patents in its search trajectory and apply the same budget~$K$ for a fair comparison.}
\small
\label{tab:retrieval_baselines}
\renewcommand{\arraystretch}{1.0}
\setlength{\tabcolsep}{20pt}
\begin{adjustbox}{max width=\linewidth}
\begin{tabular}{lcccc}
    \toprule
    \textbf{Model}
    & \textbf{R@10}
    & \textbf{R@20}
    & \textbf{R@50}
    & \textbf{R@100}\\
    \midrule
    \multicolumn{5}{
  >{\columncolor{gray!20}}c
}{
  \textbf{Sparse Retriever}
} \\
    \midrule
    TF-IDF
    & 5.49 & 7.71 &10.53 & 13.28 \\
    BM25
    & 12.84 & 15.20 & 19.68 & 23.93  \\
    uniCOIL
    & 12.37 & 15.46 &19.90 & 24.12  \\
    SPLADE-v3
    & \textbf{13.96} & \textbf{17.51} & \textbf{24.28} & \textbf{29.25}  \\
    \midrule
    \multicolumn{5}{
  >{\columncolor{gray!20}}c
}{
  \textbf{Dense Retriever}
} \\
    \midrule
    BGE-M3
    & 15.98 & 19.65 & 26.06 & 31.87  \\
    jina-embedding-v5-text-small
    & 18.08 & 23.05 & 30.32 & 37.25  \\
    jina-embedding-v5-text-nano
    & 20.73 & 25.57 & 33.40 & 39.90  \\
    Qwen3-Embedding-0.6B
    & 20.02 & 25.76 & 33.08 & 38.19 \\
    Qwen3-Embedding-4B
    & 21.66 & 27.64 & 35.62 & 42.62  \\
    Qwen3-Embedding-8B
    & 21.56 & 27.98 & 35.53 & {43.63}  \\
    PaECTER
    & 21.48 & 27.23 & 36.11 & 42.30  \\
    PatEmbed-Large
    & \textbf{22.69} & \textbf{28.60} & \textbf{38.19} & \textbf{45.98} \\
    \midrule
    \multicolumn{5}{
  >{\columncolor{gray!20}}c
}{
  \begin{tabular}{c}
    \textbf{Prior-Art Agentic Search}
    {\small\textit{w/ the Same Budget}}
  \end{tabular}
} \\
    \midrule
    Claude-Opus-5-max
    & 19.33 & \textbf{27.99} & \textbf{37.37} & \textbf{45.71} \\
    Claude-Opus-4.7-max
    & \textbf{19.89} & 27.58 & 36.00 & 40.78 \\
    Qwen3.8-Max-0813-xhigh
    & 18.26 & 27.68 & 37.74 & 40.73 \\
    GLM-5.3-max
    & 16.98 & 26.48 & 36.53 & 39.98\\
    DeepSeek-V4-Flash-0731-max
    & 17.62 & 25.19 & 36.40 & 43.25 \\
    Qwen3.7-Max-xhigh
    & 17.57 & 25.80 & 34.97 & 36.66 \\
    Qwen3.5-9B
    & 15.65 & 22.02 & 27.19 & 28.36 \\
    Qwen3.5-4B
    & 14.91 & 19.19 & 23.61 & 24.06 \\
    \multicolumn{5}{
  >{\columncolor{gray!10}}c
}{
  \textbf{\textit{Ours}}
} \\
    PatEx-Agent-4B
    & 16.10 & 24.20 & 31.80 & 36.36  \\
    \qquad\text{ }$\mathbf{\Delta}$ vs. Base & \upperformence{8.0\%} & \upperformence{26.1\%} & \upperformence{34.7\%} & \upperformence{51.1\%} \\
    \bottomrule
\end{tabular}
\end{adjustbox}
\end{table}
\paragraph{Broader prior-art discovery does not always translate into better reference selection.}
DeepSeek-V4-Flash-0731 achieves a competitive Search-R of 44.1, yet its Select-R remains at 24.7.
This discrepancy suggests that retrieving relevant references is only part of the challenge: 
agents must also identify useful documents among search results and incorporate them into the final examination.
In contrast, our PatEx-Agent improves Search-R from 23.2 to 38.7~(\upperformence{66.8\%}) while Select-R from 12.0 to 18.2 (\upperformence{51.7\%}) over its backbone.
The disproportionate gain in Select-R suggests that PatEx-Agent improves not only search coverage, but also the ability to convert retrieved evidence into selected references used in the examination.

\subsection{Prior-Art Search Performance Analysis}

\paragraph{Retrieval Performance on Patent-1M.}
As shown in Table~\ref{tab:retrieval_baselines}, we compare sparse and dense retrievers with the agents.
Dense retrievers consistently outperform sparse retrievers, with PatEmbed-Large achieving the best results at all cutoffs, surpassing the best sparse retriever, SPLADE-v3.
Notably, its advantage increases with retrieval depth, highlighting the effectiveness of domain-specific semantic
\begin{wrapfigure}{r}{0.48\linewidth}
  \centering
  \includegraphics[width=\linewidth]{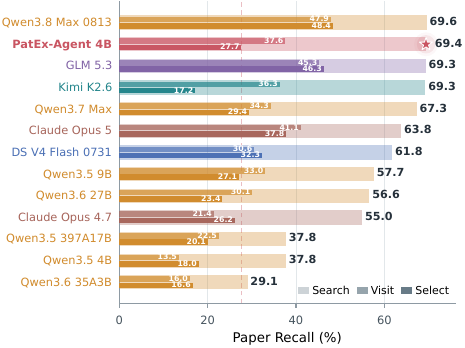}
  \caption{NPL recall in PatEx-Bench. Each bar reports NPL recall at three stages, \textit{i.e.}, Search-R, Visit-R, and Select-R.}
  \vspace{-1em}
  \label{fig:paper-search-recall-models}
\end{wrapfigure}
representations for large-scale prior-art retrieval.
Our PatEx-Agent consistently outperforms its backbone across all retrieval cutoffs, with larger gains observed at deeper ranks. Its R@100 increases from 24.06 to 36.36~(\upperformence{51.1\%}), indicating that the learned search behavior improves long-range candidate coverage. 

\paragraph{Non-Patent Literature Search Performance}
Beyond patent documents, novelty assessment in patent examination may also require searching NPL. We focus on scientific papers as a representative form of NPL, since they provide technical disclosures that can be directly compared against patents.
As shown in Figure~\ref{fig:paper-search-recall-models}, we report agent performance across the Search--Visit--Select stages for examiner-cited papers.
Our PatEx-Agent achieves 69.4 Paper Search-R, nearly matching Qwen3.8-Max at 69.6 despite its much smaller scale. Its performance remains highly competitive at later stages, reaching 37.6 Visit-R and 27.7 Select-R and outperforming many substantially larger models. 
This suggests that the learned search behavior generalizes beyond patent retrieval to NPL.

\subsection{Ablation Study}

\subsubsection{Ablation on Trajectory Sampling Strategy}
\begin{wrapfigure}{r}{0.65\linewidth}
  \centering
  \vspace{-1em}
  \includegraphics[width=\linewidth]{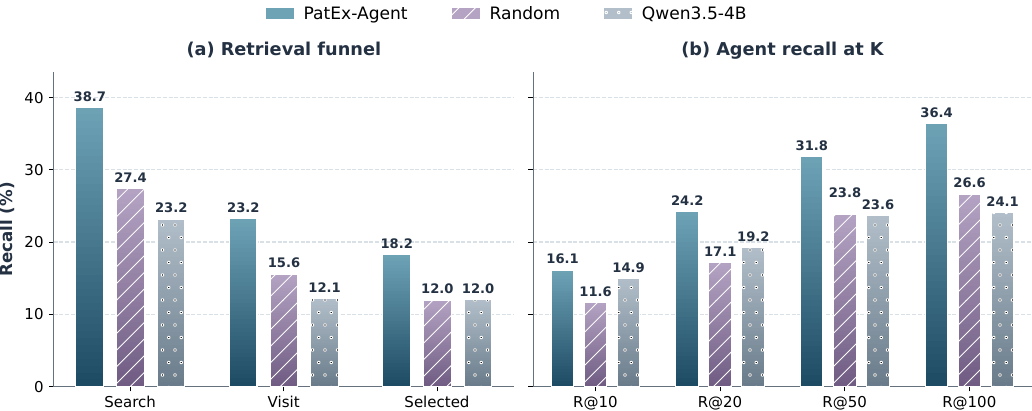}
  \caption{Ablation of sampling strategies for prior-art search.}
  \label{fig:random-ablation}
\end{wrapfigure}
To isolate the effect of trajectory selection, we randomly sample 3K trajectories from the same generated pool, matching the size of PatEx-SynData, and train the same backbone with identical settings.
Random sampling provides only limited and inconsistent gains over the Qwen3.5-4B base model. 
In particular, random sampling reduces R@10 from 14.9 to 11.6~(\downperformece{22.1\%}) and R@20 from 19.2 to 17.1~(\downperformece{10.9\%}), while Select-R also drops slightly from 12.03 to 11.96~(\downperformece{0.6\%}).
In contrast, training on PASS-filtered trajectories raises Select-R to 18.24~(\upperformence{51.6\%}). 
Under the same data budget and training procedure, this gap highlights the importance of PASS's multidimensional quality assessment helps select trajectories that teach the agent to turn retrieved candidates into useful final evidence, rather than relying on fine-tuning alone.

\subsubsection{Ablation on Search Toolset}
\begin{wrapfigure}{r}{0.65\linewidth}
  \centering
  \vspace{-1em}
  \includegraphics[width=\linewidth]{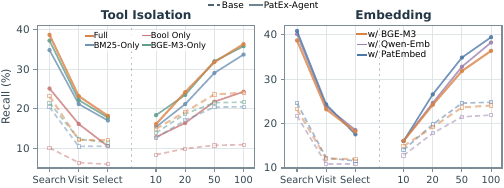}
  \caption{Retrieval recall under tool isolation (left) and semantic-retriever replacement (right).  Where Qwen-Emb refers to Qwen3-Embedding-0.6B.}
    \vspace{-1em}
  \label{fig:retrieval-tool-ablation}
\end{wrapfigure}
We conduct a search-tool ablation study by isolating the Boolean, BM25, and BGE-M3 retriever, with only one enabled.
As shown in Fig.~\ref{fig:retrieval-tool-ablation}~(left), the full toolset provides the strongest retrieval performance, while Boolean-only search causes the largest degradation. 
More importantly, under the same Boolean-only constraint, our PatEx-Agent improves Search-R from 10.08 to 25.12~(\upperformence{149.2\%}) and Visit-R from 6.35 to 16.14~(\upperformence{154.2\%}) over the backbone. 
This indicates that PatEx-Agent learns more effective query construction and search behavior from the PatEx-SynData rather than merely relying on a strong tool.
Among the isolated tools, BGE-M3 performs best and allows PatEx-Agent to nearly recover its full-tool performance in Search-R (37.24 vs. 38.70) and $R@100$ (35.82 vs. 36.36).
Nevertheless, the full toolset remains stronger in Search-R and Select-R, suggesting that Boolean, lexical, and dense retrieval provide complementary coverage.

\subsubsection{Ablation on Semantic Search Tool}
We replace the dense retriever while remains the other tools, comparing BGE-M3 with Qwen3-Embedding-0.6B and PatEmbed-Large.
As shown in Figure~\ref{fig:retrieval-tool-ablation}~(right), stronger retrievers generally improve recalls, but these gains do not always translate into better Select-R.
For Qwen3.5-4B, PatEmbed-Large increases Search-R from 23.20 to 24.60 and R@100 from 24.06 to 24.78, while Select-R slightly decreases from 12.00 to 11.60.
This may indicate that stronger retriever introduces more similar candidates, making evidence selection more difficult.

\subsection{Case Study}
\label{sec:case-study}


\begin{figure*}[!t]
\centering

\begingroup

\definecolor{trajHeader}{HTML}{454A4D}
\definecolor{trajInk}{HTML}{182330}
\definecolor{trajMuted}{HTML}{69727A}
\definecolor{trajApplication}{HTML}{F1ECFA}
\definecolor{trajBlueBg}{HTML}{EAF3FA}
\definecolor{trajGray}{HTML}{F3F4F4}
\definecolor{trajGreenBg}{HTML}{EAF5F1}
\definecolor{trajPurple}{HTML}{F2EDFA}

\definecolor{trajRed}{HTML}{C84F5C}
\definecolor{trajBlue}{HTML}{3E779D}
\definecolor{trajOrange}{HTML}{A9632F}
\definecolor{trajGreen}{HTML}{307765}

\newcommand{\trajrow}[4]{%
  \begin{tcolorbox}[
    colback=#1,
    colframe=#1,
    boxrule=0pt,
    arc=.25mm,
    left=.70mm,
    right=.70mm,
    top=.15mm,
    bottom=.15mm,
    before skip=.12mm,
    after skip=0mm,
    equal height group=#4
  ]
    \fontsize{6.4}{6.8}\selectfont
    \color{trajInk}
    #2\hspace{.25em}#3
  \end{tcolorbox}%
}

\newcommand{\trajapp}[3]{%
  \begin{tcolorbox}[
    colback=trajApplication,
    colframe=trajApplication,
    boxrule=0pt,
    arc=.30mm,
    left=.70mm,
    right=.70mm,
    top=.22mm,
    bottom=.22mm,
    before skip=0mm,
    after skip=.12mm,
    equal height group=#3
  ]
    \fontsize{6.4}{6.8}\selectfont
    \color{trajInk}
    \textbf{Patent Application #1}\\[-.1mm]
    #2
  \end{tcolorbox}%
}

\newcommand{\trajcompare}[4]{%
  \begin{tcolorbox}[
    colback=trajGray,
    colframe=trajGray,
    boxrule=0pt,
    arc=.25mm,
    left=.70mm,
    right=.70mm,
    top=.22mm,
    bottom=.22mm,
    before skip=.15mm,
    after skip=0mm,
    equal height group=#4
  ]
    \fontsize{6.2}{6.5}\selectfont
    \color{trajInk}

    \textbf{\textcolor{trajMuted}{Ground Truth:}}
    \hspace{.35em}#1\\[-.1mm]

    \textbf{\textcolor{trajGreen}{Model Output:}}
    \hspace{.35em}#2\\[-.1mm]

    #3
  \end{tcolorbox}%
}

\begin{minipage}[t]{0.492\textwidth}
\vspace{0pt}

\begin{tcolorbox}[
  width=\linewidth,
  enhanced,
  colback=white,
  colframe=trajHeader,
  boxrule=.45pt,
  arc=.8mm,
  left=1.0mm,
  right=1.0mm,
  top=.8mm,
  bottom=.8mm,
  colbacktitle=trajHeader,
  coltitle=white,
  title={Example I: Relevant Prior Art Retained},
  fonttitle=\bfseries\footnotesize,
  toptitle=.35mm,
  bottomtitle=.35mm
]

\trajapp
  {17/719,110}
  {Claim 1: request teleoperator guidance for candidate actions,
   then update the control model to increase autonomy.}
  {traj-app}

\trajrow
  {trajGray}
  {\textbf{[Think]:}}
  {Search for assistance triggers and feedback-based learning.}
  {traj-think}

\trajrow
  {trajBlueBg}
  {\textbf{\textcolor{trajBlue}{[Search]:}}}
  {Patent BM25: ``robot candidate actions teleoperation''}
  {traj-search-1}

\trajrow
  {trajGray}
  {\textbf{[Result]:}}
  {\textbf{\textcolor{trajBlue}{US20210031364A1 (Groz)}}---
   teleoperation fallback and model-update data.}
  {traj-result-1}

\trajrow
  {trajBlueBg}
  {\textbf{\textcolor{trajBlue}{[Search]:}}}
  {Semantic: ``candidate actions, success scores, remote assistance''}
  {traj-search-2}

\trajrow
  {trajGray}
  {\textbf{[Result]:}}
  {\textbf{\textcolor{trajRed}{US20230168670A1 (Reventlow)}}---
   candidate scoring and assistance requests.}
  {traj-result-2}

\trajrow
  {trajGreenBg}
  {\textbf{\textcolor{trajGreen}{[Visit]:}}}
  {Inspect both references' claims and descriptions.}
  {traj-visit}

\trajrow
  {trajGray}
  {\textbf{[Evidence]:}}
  {von Reventlow: action ranking; Groz:
   feedback-driven model updates.}
  {traj-evidence}

\trajrow
  {trajGreenBg}
  {\textbf{[Select]:}}
  {von Reventlow (primary) + Groz (secondary).}
  {traj-select}

\trajrow
  {trajPurple}
  {\textbf{\textcolor{trajRed}{[Office Action]:}}}
  {Reject claims 1--20 under 35 U.S.C.~\S~103.}
  {traj-oa}

\trajcompare
  {Non-Final Rejection $\mid$ \S~103}
  {Non-Final Rejection $\mid$ \S~103}
  {\textbf{\textcolor{trajOrange}{Reference Recall:}}
   Search 100 $\mid$ Visit 100 $\mid$ Select 100}
  {traj-compare}

\end{tcolorbox}

\end{minipage}%
\hfill%
\begin{minipage}[t]{0.492\textwidth}
\vspace{0pt}

\begin{tcolorbox}[
  width=\linewidth,
  enhanced,
  colback=white,
  colframe=trajHeader,
  boxrule=.45pt,
  arc=.8mm,
  left=1.0mm,
  right=1.0mm,
  top=.8mm,
  bottom=.8mm,
  colbacktitle=trajHeader,
  coltitle=white,
  title={Example II: Retrieved but Not Selected},
  fonttitle=\bfseries\footnotesize,
  toptitle=.35mm,
  bottomtitle=.35mm
]

\trajapp
  {19/452,413}
  {Claim 1: verify an FHE-encrypted machine-learning model
   through a smart contract and conditionally transfer cryptocurrency.}
  {traj-app}

\trajrow
  {trajGray}
  {\textbf{[Think]:}}
  {Decompose FHE verification, contract logic, and payment control.}
  {traj-think}

\trajrow
  {trajBlueBg}
  {\textbf{\textcolor{trajBlue}{[Search]:}}}
  {Patent BM25: ``FHE smart contract ML verification''}
  {traj-search-1}

\trajrow
  {trajGray}
  {\textbf{[Result]:}}
  {\textbf{\textcolor{trajBlue}{US20190334716A1 (Kocsis)}}---
   encrypted model testing and transfer criteria.}
  {traj-result-1}

\trajrow
  {trajBlueBg}
  {\textbf{\textcolor{trajBlue}{[Search]:}}}
  {Paper BM25: ``smart contract ML verification FHE escrow''}
  {traj-search-2}

\trajrow
  {trajGray}
  {\textbf{[Information]:}}
  {\textbf{\textcolor{trajRed}{smartFHE}}
   (GT, rank 9)---retrieved, but not visited or selected.}
  {traj-result-2}

\trajrow
  {trajGreenBg}
  {\textbf{\textcolor{trajGreen}{[Visit]:}}}
  {Inspect Kocsis, Rosen, and the Guo paper.}
  {traj-visit}

\trajrow
  {trajGray}
  {\textbf{[Evidence]:}}
  {Kocsis: verification; Guo: encrypted features;
   Rosen: conditional payment.}
  {traj-evidence}

\trajrow
  {trajGreenBg}
  {\textbf{[Select]:}}
  {Kocsis + Rosen + Guo et al. (alternative evidence).}
  {traj-select}

\trajrow
  {trajPurple}
  {\textbf{\textcolor{trajRed}{[Office Action]:}}}
  {Reject claims 1--2 under \S~103 over Kocsis, Guo, and Rosen.}
  {traj-oa}

\trajcompare
  {Non-Final Rejection $\mid$ \S\S~101, 103}
  {Non-Final Rejection $\mid$ \S~103}
  {\textbf{\textcolor{trajOrange}{Reference Recall:}}
   Search 100 $\mid$ Visit 0 $\mid$ Select 0}
  {traj-compare}

\end{tcolorbox}

\end{minipage}

\endgroup

\caption{
Case study of two AI-domain patent applications.
The left example retains relevant prior art through search, inspection, and selection,
while the right example retrieves the relevant reference but fails to use it in the final evidence set.
}
\label{fig:prior-art-trajectory-example}

\end{figure*}

We provide a detailed case study with two contrasting examples in Figure~\ref{fig:prior-art-trajectory-example}, covering successful prior-art retention and failure to use retrieved evidence in the final decision.
\textbf{(I)} In the left example, we observe a successful pipeline: the agent decomposes the claim into complementary concepts, reformulates its query across retrieval tools, verifies candidate prior-art, and assigns distinct evidentiary roles before generating the OA. 
It preserves the oracle reference throughout the Search--Visit--Select funnel and yields the correct decision and statutory ground.
\textbf{(II)} In the right example, we observe a different pattern in which the agent retrieves the oracle reference but does not promote it to document inspection or final selection. 
Its statutory prediction is incomplete but does not introduce incorrect grounds.
Together, these examples show that effective agentic examination requires both iterative search and disciplined evidence selection: 
broad retrieval can expose the necessary prior art, but only verified and correctly selected novelty-aware evidence can support a complete legal conclusion.

\section{Conclusion}
In this work, we introduced \textbf{PatEx-Gym}, a unified framework for developing and evaluating agentic patent examination models, including \textbf{PatEx-Bench}, {PASS} for synthesizing \textbf{PatEx-SynData}, and the lightweight baseline \textbf{PatEx-Agent}. 
Our extensive experiments show that final decision accuracy alone cannot fully capture examination quality, and that prior-art discovery, evidence inspection, and final selection remain challenging even for strong frontier models.
Despite its 4B scale, our PatEx-Agent substantially improves over its base model and achieves prior-art discovery performance comparable to agents with \(\sim\)30B size. 
These findings identify evidence inspection and selection as central challenges for agentic search and deep research. 
Moreover, we believe that effective agents must go beyond document search to coordinate adaptive querying, tool-use, evidence integration, and reasoning throughout the search process. 
We hope our PatEx-Gym provides a realistic testbed for advancing evidence-grounded agentic search and deep research.

\bibliography{ref}

\appendix

\end{document}